\documentclass[a4paper]{IEEEtran}
\usepackage[numbers]{natbib}
\usepackage{comment}
\usepackage{graphicx}
\usepackage[numbers]{natbib}
\usepackage{mathdots}
\usepackage{float}

\usepackage{authblk}

\usepackage{fancyhdr}

\fancypagestyle{plain}{%
\fancyhf{}%
\fancyfoot[L]{\footnotesize \center \emph{Preprint accepted for publication in the Proceedings of the IEEE International Conference on Signal Processing and Communications (SPCOM), 2020.  \\ -\thepage-}}%
}

\begin{document}

\pagestyle{fancy}
\fancyfoot[C]{Preprint accepted for publication in the Proceedings of the IEEE International Conference on Signal Processing and Communications (SPCOM), 2020.}
\title{Acoustic prediction of flowrate: varying liquid jet stream onto a free surface}
%
%


\author[1]{Balamurali B T}
\author[2]{Edwin Jonathan Aslim}
\author[2]{Yun Shu Lynn Ng}
\author[2]{Tricia Li Chuen Kuo}
\author[1]{Jacob Shihang Chen}
\author[1]{Dorien Herremans}
\author[2]{Lay Guat Ng}
\author[1]{Jer-Ming Chen}

\affil[1]{Singapore University of Technology and Design, Singapore 487372 \authorcr Email: {\tt \{balamurali\_bt, dorien\_herremans, jerming\_chen\}@sutd.edu.sg}\vspace{1.5ex}}
\affil[2]{Singapore General Hospital, Singapore 169856 \authorcr  \vspace{-2ex}} 


%
%
\pagestyle{fancy}
\maketitle              
\pagestyle{fancy}
\begin{abstract}
Information on liquid jet stream flow is crucial in many real world applications. In a large number of cases, these flows fall directly onto free surfaces (e.g. pools), creating a splash with accompanying splashing sounds. The sound produced is supplied by energy interactions between the liquid jet stream and the passive free surface. In this investigation, we collect the sound of a water jet of varying flowrate falling into a pool of water, and use this sound to predict the flowrate and flowrate trajectory involved. Two approaches are employed: one uses machine-learning models trained using audio features extracted from the collected sound to predict the flowrate (and subsequently the flowrate trajectory). In contrast, the second method directly uses acoustic parameters related to the spectral energy of the liquid-liquid interaction to estimate the flowrate trajectory. The actual flowrate, however, is determined directly using a gravimetric method: tracking the change in mass of the pooling liquid over time. We show here that the two methods agree well with the actual flowrate and offer comparable performance in accurately predicting the flowrate trajectory, and accordingly offer insights for potential real-life applications using sound.

\begin{keywords}
Flowrate Trajectory, Flowrate, Laminar Stream, Audio Features, Machine Learning
\end{keywords}

\end{abstract}
\section{Introduction}
The flow of falling liquids is found in many real-world applications (e.g. waterfalls, flushing of toilets, rain, weirs, fountains, pump exhausts, etc.). Information on the manner of its flow, flowrate and its dynamics offers crucial insights into these phenomena and the systems accompanying them. In a large number of cases, liquid streams may fall directly onto free surfaces (i.e., pools) and against solid walls, resulting in splashing sounds associated with the stream interacting with the passive free surface, via energy and fluid dynamic interactions (e.g. momentum of stream/droplets, surface tension, viscosity etc).

While detailed analyses of the dynamic regimes of such liquid-liquid interactions are complicated, it has been widely observed that the sound produced by these interactions can reveal key aspects of the flow interactions and other associated physical properties of the flow system. This insight makes sense, as different combinations of fluid dynamics result in distinct gas-liquid interactions that create distinct acoustic responses which can serve as acoustic signatures \cite{aabom2006introduction}.

In \cite{diatschenko1994passive}, the flow regime type of a gas-liquid mixture flowing through a pipe is determined by estimating the acoustic spectrum of the flow mixture passing through the conduit and comparing it against the acoustic fingerprints associated with various flow regimes. Similarly, in \cite{belchamber1992method}, the characteristics of sluggish flow (frequency of slugging, velocity and length of slugs) in a multiphase flow pipeline is determined from the acoustic emission in the ultrasonic frequency range. In \cite{sustek1980steam}, the continuous steam quality is determined from the acoustics of the steam output through an orifice (sonic vibrations created by the steam flow). A signal proportional to the quality of the steam is created from the acoustic interaction by filtering the sonic vibrations for a particular narrowband frequency. In \cite{diatschenko1998passive}, acoustic techniques are used to analyse noise generated by fluid flow within a pipe. This is then used to measure the internal diameter of the pipe. In \cite{diatschenko1995acoustic}, physical properties of fluid flow within a pipe such as various mass flowrates, the void-to-liquid fraction, fluid density, and velocity are determined by measuring the characteristic acoustic frequency of the pipe and its amplitude variation associated with differential pressure. In the field of healthcare, there are studies that determine the urinary flowrate trajectory by analyzing characteristic sounds (spectral data) using regression methods, which are then applied in monitoring contexts to indicate different flow patterns \cite{belotserkovsky2013urine,brohan2010systems,song2018comparison,aslim2019novel}.



In this report, the sound of varying water streams falling onto a pool of water is used to predict the flowrate and flowrate trajectory of the streams. Two different approaches are developed and investigated here: the first method uses machine-learning (ML) models that are trained using audio features extracted from the collected sound to predict the flowrate and the flowrate trajectory. The second method does not involve any machine learning part, but instead directly uses acoustic parameters related to the spectral energy of the liquid interaction to estimate the flowrate trajectory (cf. Fig.~\ref{flowpattern}) in question. Consequently, the efficacy of both these methods are compared and their pros and cons discussed. 

The remainder of this paper is organized as follows: Section \ref{Sec_2} (Methodology) offers a brief overview of the used hardware and data collection protocol, flowrate trajectory extraction procedures, and an overview of various distance measures used for measuring the efficacy of the investigated procedures. Details of experimental results are presented in Section \ref{Sec:Expresults}. Finally, conclusions from this investigation is presented at Section \ref{Sec:conclu}.

\section{Methodology}
\label{Sec_2}
\subsection{Flowrate Data Collection}
Using the microphone on a smartphone (iPhone 5) located 1 m away, in an acoustically untreated room, we collected  audio recordings (sampling rate of 44.1 kHz) of water stream having varying flowrate, together with its flowrate measured using a gravimetric flowmeter (LABORIE, Mississauga, ON, Canada). A varying water jet stream coming out of a narrow tube was directed into the flowmeter. The flowmeter consists of a collecting funnel and a beaker placed underneath, on a digital weighing scale connected to a computer.  This scale thus reads the change in the beaker's weight over time (every 0.1s) as the liquid stream collects into it, thereby indicating flowrate (resolution of 1ml/0.1s).  Acoustically, we treat the entire interaction between the water jet and the flowmeter as one system (``blackbox"). A total of 51 flow episodes (i.e., 51 flowrate readings) along with their audio recordings were collected. It has to be noted that the collecting beaker is rather wide and deep (volume of almost a thousand of millilitres of air), and the flowrate of water stream entering the container is relatively small (several millilitres of water per second), consequently, for the flow period observed (several seconds), the Helmholtz volume is effectively constant (the change in water level is negligible); any Helmholtz resonance associated with the air in the container is effectively unvarying and so does not give rise to any spurious acoustic artefacts or statistical generalizations, associated with our data collection.


\subsection {Flowrate Trajectory Extraction}
\subsubsection{Machine Learning Method}\ \

\begin{figure}
\includegraphics[scale = 0.22]{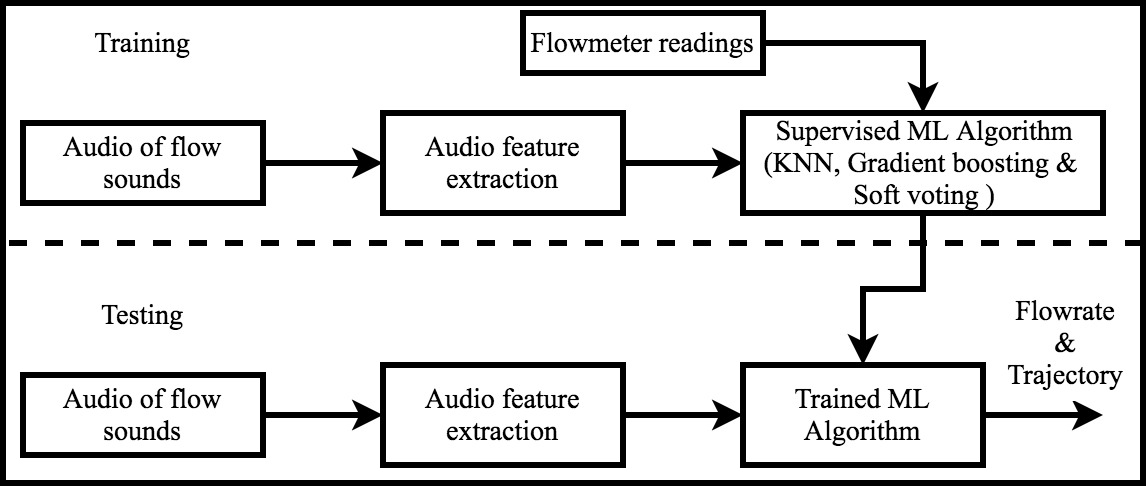}
\caption{Flowrate trajectory prediction using machine learning method.} \label{ML_methodology}
\end{figure}

Each audio recording, paired with the corresponding flowmeter reading, were used to train the machine-learning algorithm (Flowrate is determined from the change of liquid volume collected over time). The methodology used in this investigation is shown in Fig.~\ref{ML_methodology}.

To synchronize with the collected flowrate measurements, audio recordings were similarly divided into 0.1 second frames. From every frame, 14 Mel-frequency Cepstral Coefficients (MFCCs) were extracted. MFCCs were chosen because of their ubiquity in the audio recognition arena \cite{rabiner1999fundamentals,rubin2016classifying,lee2008automatic,piczak2015environmental,balamurali2019towards}. Two supervised machine learning algorithms: k-nearest neighbours (KNN) \cite{keller1985fuzzy} and gradient boosting \cite{friedman2002stochastic}, were trained using the extracted MFCCs and their corresponding flowmeter labels. The number of nearest neighbours was set to 200 and all of the points in each neighbourhood were weighted equally.  Minkowski distance was chosen as the distance metric (with a power parameter equal 2). For the gradient boosting, 100 boosting stages were used with a learning rate of 0.1. The quality of a tree split was based on mean squared error.  

From the dataset collected, 70\% of the recordings were used for training the ML algorithm, i.e., 35 recordings which resulted in a total of more than 10000 audio frames of 0.1 second. The remaining 30\% was used for testing, i.e., 16 recordings (4,500 audio frames) \cite{crowther2005method}. To test the algorithm, the MFCCs extracted from every audio frame of the testing set were passed into the pre-trained classifiers to predict the corresponding volume. Final volume was estimated by soft-voting the predictions  from the two classifiers (soft voting returns the predicted label as argmax of the average of predicted probabilities from two classifiers). Resulting flowrate trajectories were derived from this estimated volume. In order to aid comparison with acoustic parameter method (described below), the resulting flowrate trajectory was normalized with a min-max normalization with minimum zero and maximum one.

\subsubsection{Acoustic Parameter Method}\ \

\begin{figure}
\includegraphics[scale = 0.25]{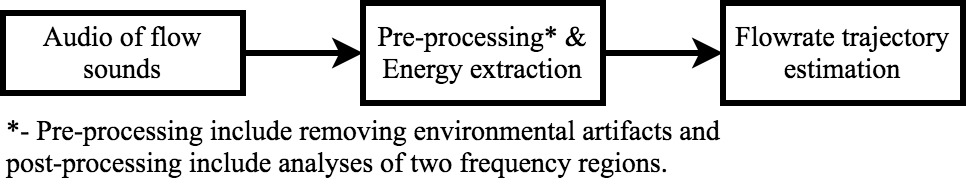}
\caption{Flowrate trajectory estimation using acoustic parameter method.} \label{Energy_contour}

\end{figure}

The dynamic interaction of a liquid stream entering a free surface is complex (quasi-chaotic) in nature, coupled with the fact that under certain flow regimes, the stream may in fact break up mid-stream to coalesce as a sequence of droplets; the consequent impulsive interactions with the free surface results in complicated gas-liquid surface interactions (``splashing'' and ``bubbling'') and even secondary splashes. Nevertheless, this spectrum of complex interactions consistently produce sustained broadband sounds with distinct acoustic parameters which are directly related with the energy of the interaction (cf. conservation of momentum) and may be meaningfully exploited to reveal and estimate the flowrate trajectory. The corresponding methodology is shown in Fig.~\ref{Energy_contour}. 

The raw audio recordings were found to contain artifacts introduced by the environment (e.g. air conditioning noise) and required pre-processing to account for this. 
Further, inspection of the spectrogram (visually) showed energy concentrated in two distinct frequency regions which may be meaningfully related to the flowrate: a low frequency region ($<$ 1 kHz) and a high frequency region ($>$ 1 kHz). When the audio samples were listened, the low frequency region seemed to associate with direct impact of the stream  while the high frequency region seemed to associate with secondary splashes.    

\begin{figure}
\includegraphics[scale = 0.30]{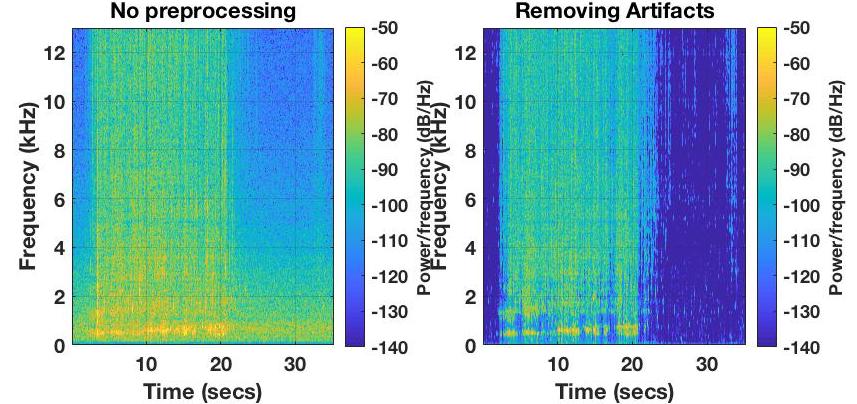}
\caption{Spectrogram of an audio recording before and after pre-processing.} \label{Spectrogram1}
\end{figure}

For pre-processing the audio recordings, harmonic regeneration noise reduction (HRNR) technique was used to account for the artifacts introduced by the room and the background noise \cite{plapous2006improved}. This technique produces good results in removing noise. As a result it was found to improve the perceptual quality of the audio recording (Fig.~\ref{Spectrogram1} shows spectrogram of an audio recording before and after pre-processing using HRNR technique.).

After pre-processing, the energy associated with the audio sample was extracted for each audio frames. These values were smoothed to obtain (na\"ively) the flowrate trajectory (flowrate trajectory was again min-max normalized). 

Further, we extracted energy from the two frequency regions identified earlier to see if the predicted flowrate trajectory could be further improved. We tried four different approaches. In first, we limited our analysis only to the low frequency part. In the second,  only the high frequency region was analysed.  Thirdly, we boosted the low frequency regions with a gain of 12 dB and examined the resulting flowrate trajectory. A similar investigation was done for the high frequency region in the fourth approach (No filtering was done in third and fourth approach).

\subsection {Measure of closeness between flowrate trajectories} 
The closeness between the predicted flowrate trajectories and the actual flowrate trajectories is reflected by the distance between them. Two distance measures were used in this investigation: Euclidean distance and Fr\'eschet distance. The lower the estimated distance the closer the predicted trajectory to the actual.
\subsubsection{Euclidean Distance}
The straight-line distance between two points is given by the Euclidean Distance \cite{anton2013elementary}.  We represent the predicted flowrate corresponding to $n$ points in a predicted flowrate trajectory (P) as ${(p_1, p_2,..., p_n)}$,
and the actual flowrate corresponding to $n$ points in actual flowrate trajectory (A) by  ${(a_1, a_2, ..., a_n)}$. The Euclidean distance between these two flowrate trajectories ${E(P,A)}$ is then given as
\begin{equation}
E(P,A) = \sqrt{\sum_{i=1}^n (p_i - a_i)^2}
\end{equation}

\subsubsection{Fr\'eschet Distance}
Fr\'eschet distance estimates the similarity between curves by taking into account the location and ordering of the points along the curves \cite{eiter1994computing}. In this investigation, a discrete variation of Fr\'eschet distance, namely coupling distance $\delta_{df}$ (i.e., looking at all the possible coupling between points from the curves), is used. The  discrete Fr\'eschet distance $\delta_{df}(P,A)$ between the predicted flowrate trajectory (P) and  the actual flowrate trajectory (A) is given as
\begin{equation}
\delta_{df}(P,A) = min{||L||}
\end{equation}
where $L$ is the coupling between P and A and $||L||$ is the longest link in L. For a detailed description of Fr\'eschet distance, see \cite{eiter1994computing}.

\section{Experimental Results}
\label{Sec:Expresults}
\subsubsection{Flowrate Trajectory Prediction}
\begin{figure}
\centering
\includegraphics[scale = 0.24]{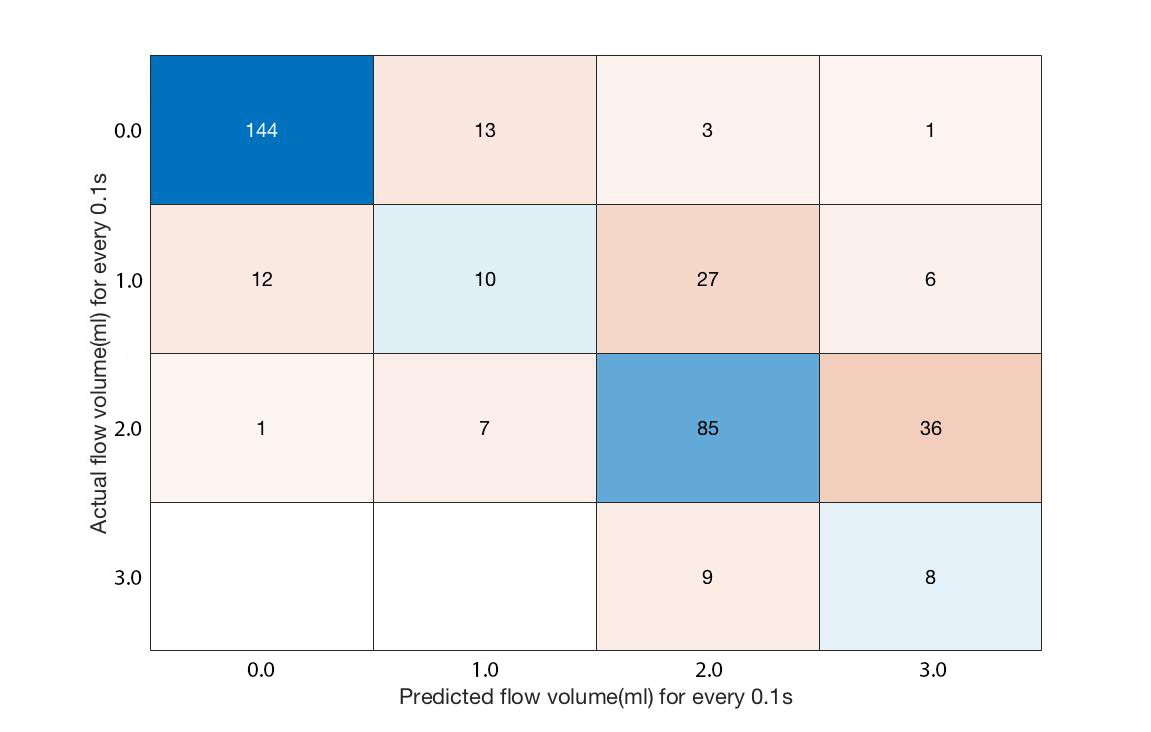}
\caption{An example confusion matrix showing accuracy of softvoting classifier for a particular recording (corresponding to that of recording ``A''). The numbers shown in each of the squares is the number of audio frames classified accordingly.} \label{confusionmatrix}
\end{figure}

\begin{figure}
\centering
\includegraphics[scale = 0.16]{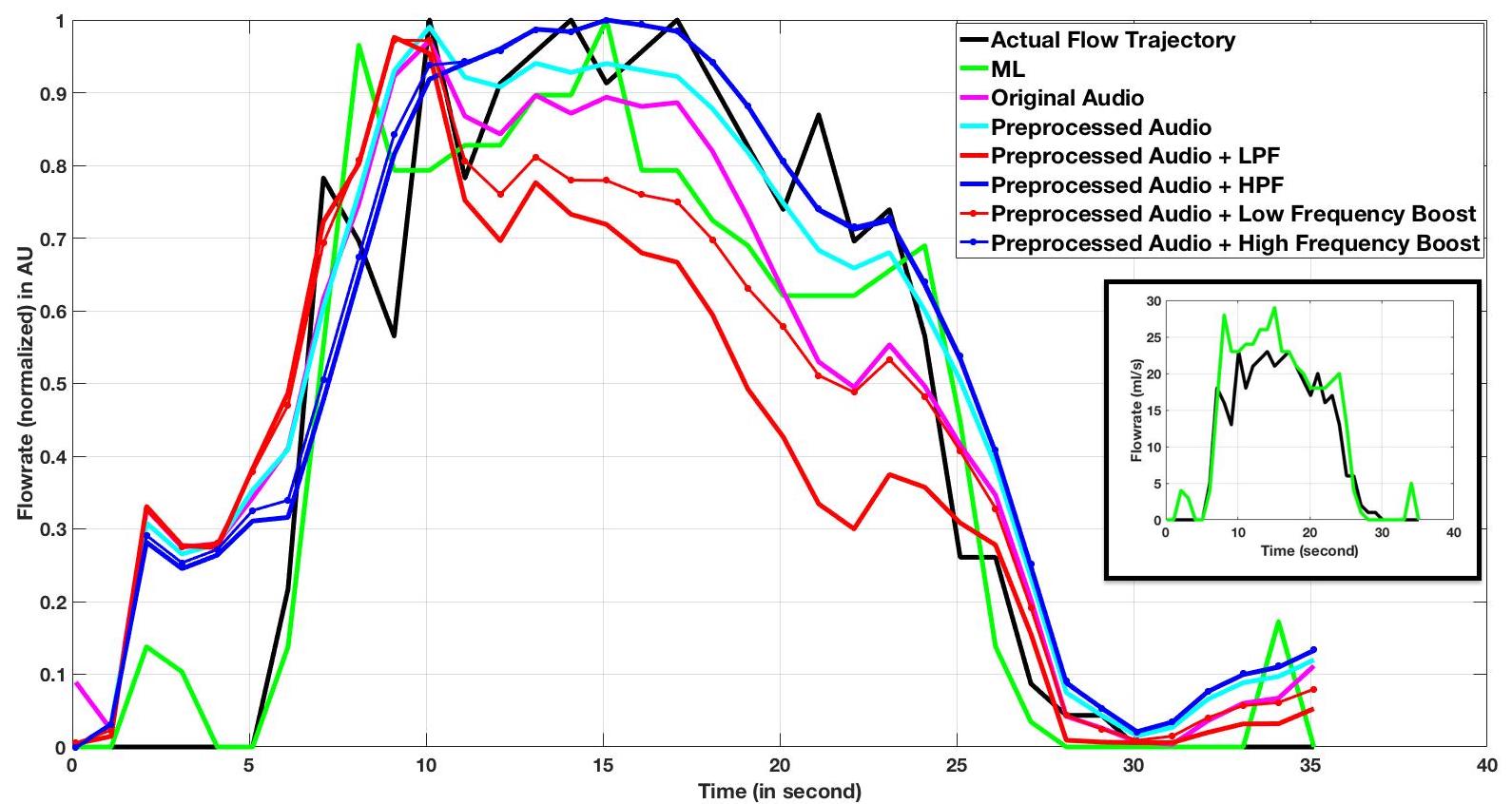}
\caption{Flowrate trajectories (corresponding to that of recording ``A'') resulting from the ML method and acoustic parameter method. To facilitate meaningful comparison,  the main figure shows the normalized result; the inset shows the  non-normalized result produced using ML method compared with the actual flowrate.} \label{flowpattern}
\end{figure}

MFCCs extracted from every audio frame of the testing recordings were used to predict
the corresponding volume using each of the trained classifiers. The resulting flow volume corresponding to the audio frame in question is estimated by soft-voting the predictions from the two classifiers mentioned earlier. As an example Fig.~\ref{confusionmatrix} presents the resulting confusion matrix and  shows the predicted and actual volume values associated with audio frames corresponding to a typical recording (that of flow recording ``A''). Confusion matrices are used to assess the performance of classifiers; for a good classifier, the resulting confusion matrix will look dominantly diagonal (i.e., showing high number of correct classification); all the off-diagonal elements represent the misclassified data. As seen in  Fig.~\ref{confusionmatrix}, our ML algorithm successfully predicts when the flow volume associated with an audio frame is 0.0 ml and 2.0 ml. However, when the actual volume  is 1.0 ml, our algorithm tends to overestimate. Consequently for this example  (see inset of Fig.~\ref{flowpattern}), our algorithm slightly overestimate the flowrate.

Fig.~\ref{flowpattern} shows the resulting flowrate trajectory predicted (Normalized flowrate trajectory shown in main part of Fig.~\ref{flowpattern} and the inset shows the non-normalized result). Also shown in Fig.~\ref{flowpattern} are the flowrate trajectories estimated for the particular flow episode using the acoustic parameter method. It is important to note that the acoustic parameter method is unable to predict volume corresponding to an audio frame, therefore a confusion matrix cannot be created for this method.

\begin{figure}[h!]
\includegraphics[scale = 0.16]{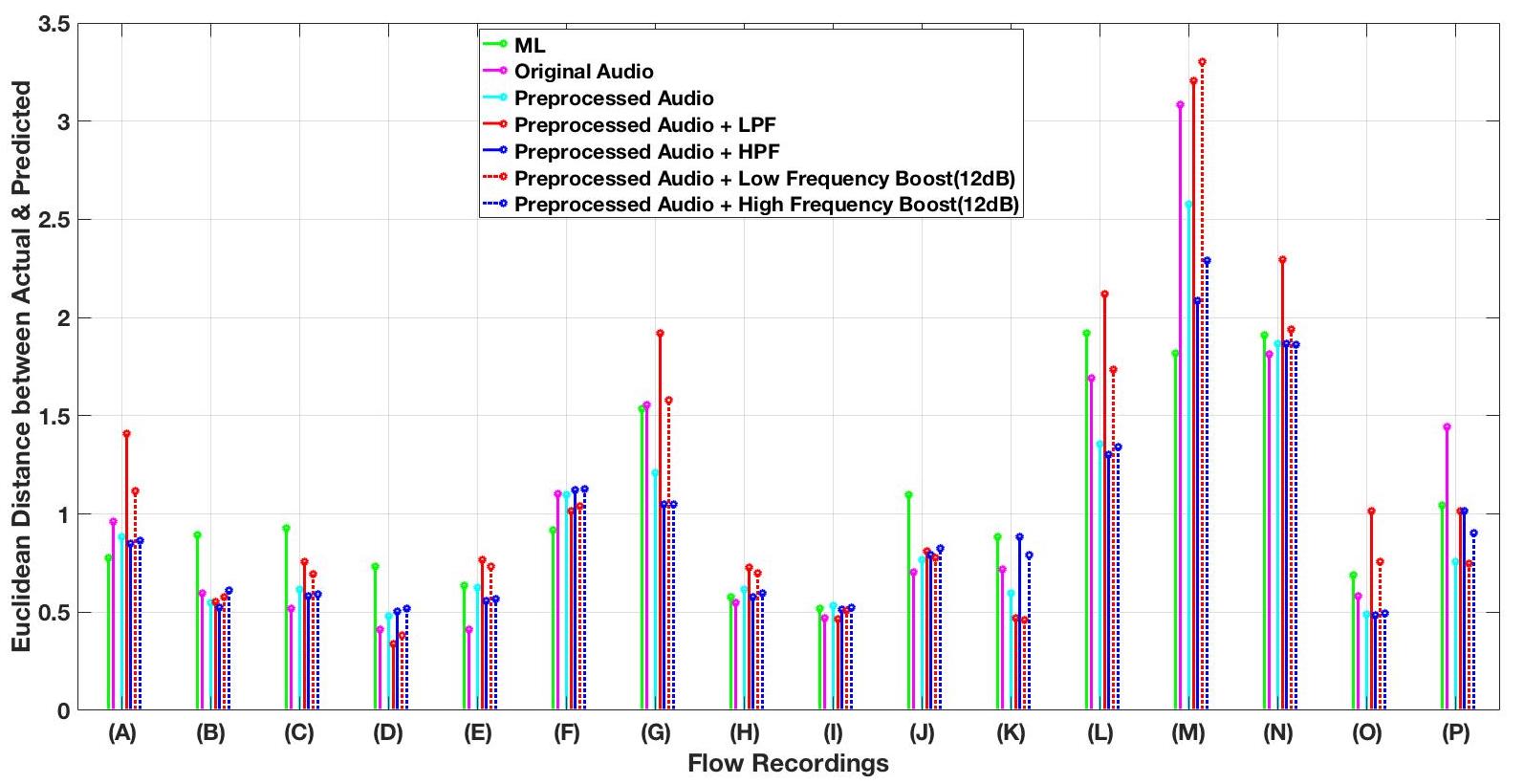}
\caption{Euclidean Distance between flowrate trajectories from prediction and actual.} \label{Euclidean_pattern}
\end{figure}

\begin{figure}[h!]
\includegraphics[scale = 0.16]{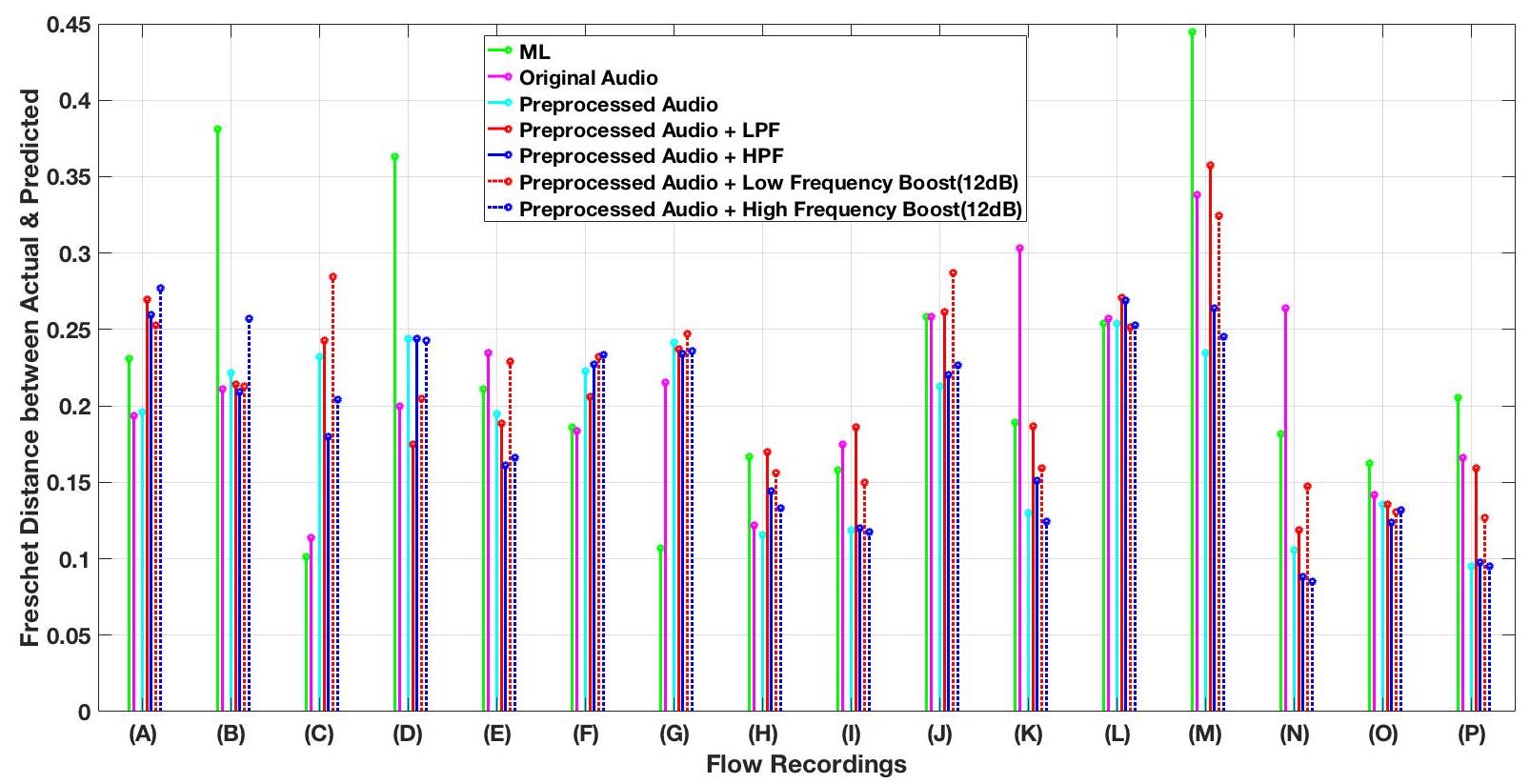}
\caption{Fr\'eschet Distance between flowrate trajectories from prediction and actual} \label{Freschet_pattern}
\end{figure}

At the onset region of the Fig.~\ref{flowpattern} (specifically from 0 to 5 s), the flowrate trajectories estimated using the acoustics parameter method were found to overestimate the flowrate. Among the two distinct frequency regions analyzed, flowrate trajectories resulting from the low frequency region  underestimates the flowrate (see the solid red curve) whereas the trajectory resulting from the high frequency band tends to overestimate the flowrate (see the solid blue curve). The low frequency boosting slightly improved the result while the high frequency boosting did not have much effect on the resulting trajectory. The closeness of the predicted trajectories to the original trajectories is compared by calculating the Euclidean distance and Fr\'eschet distance  between them. The corresponding results are shown in  Fig.~\ref{Euclidean_pattern} and  Fig.~\ref{Freschet_pattern}, respectively. Distances between the trajectories was determined for all 16 recordings in the testing set (represented as A, B, ..., P in Fig.~\ref{Euclidean_pattern} and  Fig.~\ref{Freschet_pattern}). The flowrate trajectory shown in Fig.~\ref{flowpattern} is that of recording A. 

When looking at the straight line distance between the points in flowrate trajectories of the actual and predicted cases, the
performance of our proposed ML method and the acoustic parameter approach is comparable. The ML prediction has a slight edge over the acoustic parameter approach for cases such as A, F and M, which is reflected as the lowest Euclidean distance for these trajectories as shown in  Fig.~\ref{Euclidean_pattern}. Among our various acoustic parameter approaches, the trajectories predicted using the low frequency band of spectrum were found to underestimate the relative flowrate,  and results in the worst trajectory reconstruction (i.e., comparatively high Euclidean distance). Boosting the low frequency power minimizes this trend somewhat but not by too much, whereas, trajectory reconstruction using the high frequency band of the spectrum is comparable to that of ML. 

If the similarity between trajectories is estimated by calculating the Fr\'eschet distance (i.e., accounting for location and ordering of the points in the trajectories), there is no clear distinction between the performance of the ML and acoustic parameter method (See Fig.~\ref{Freschet_pattern}). Whether this result is an artifact introduced by the normalization applied requires further investigation. 

\subsubsection{Flow Volume Prediction}
\begin{figure}[h]
\includegraphics[scale = 0.16]{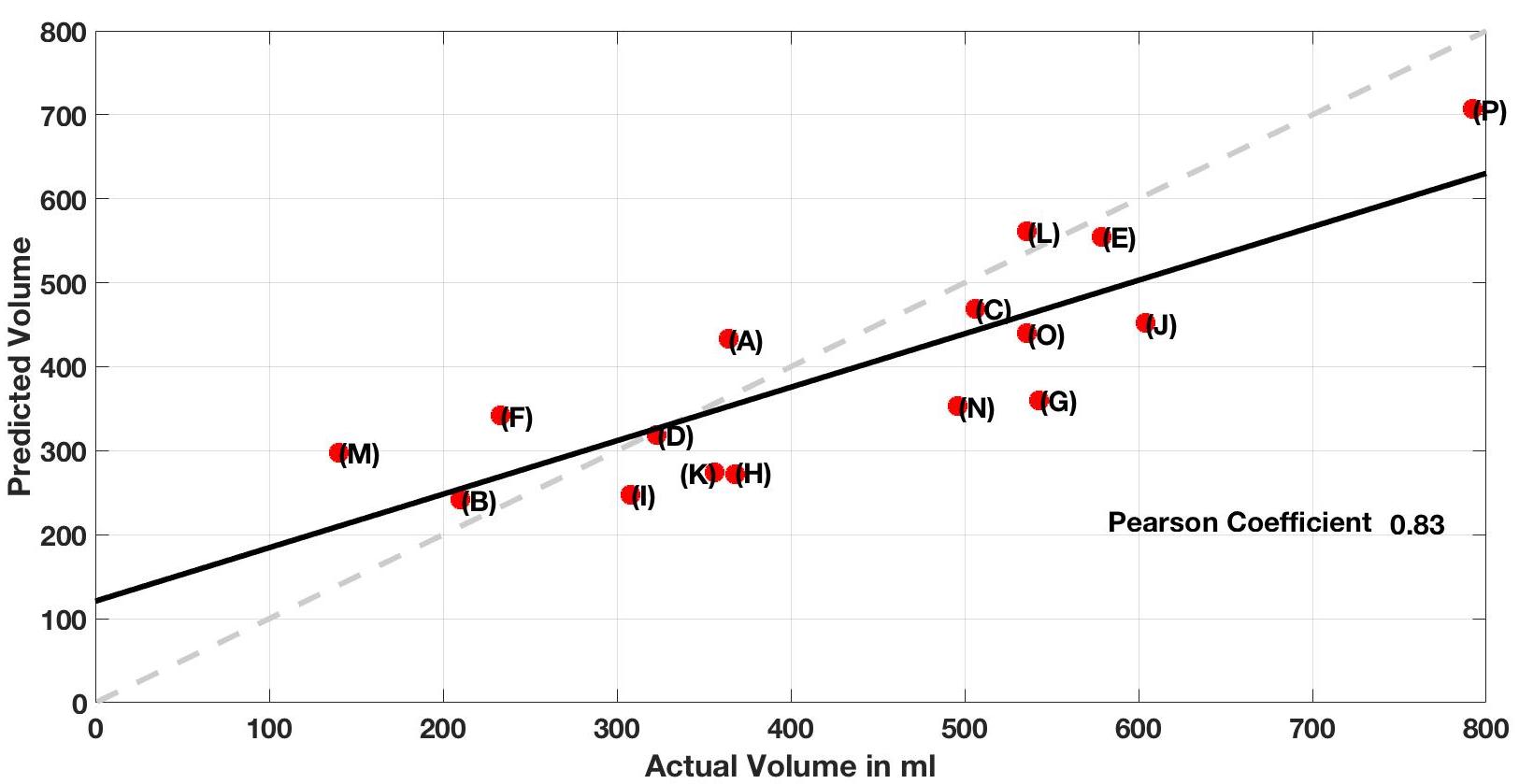}
\caption{Flow volume prediction using machine learning. The best fit line for the predicted volume is shown as a solid black line (pearson correlation coefficient 0.83) while the line corresponding to a perfect prediction is the dashed grey line (predicted volume = actual volume; (pearson correlation coefficient 1.0).} \label{Flowvolume}
\end{figure}

The ML method for flowrate prediction has inherent overheads such as the requirement of having a large number of training samples, and requires lengthy computing time to extract audio features and to train the model compared to acoustic parameter method, the latter yields only \textit{relative} flowrate. However, our ML method offers a way to directly predict the \textit{absolute} flowrate for every audio frame. Using this absolute flowrate information, not only can we predict the flowrate trajectory, then also thus determine the absolute volume associated with every recording. The predicted flow volume is found to have high correlation to the actual volume (Pearson correlation coefficient of 0.83). This result is shown in Fig.~\ref{Flowvolume}. Consequently, the best-fit line (Fig.~\ref{Flowvolume}) may be further be exploited for calibration of such an approach yielding even more accurate flowrate trajectories.

\section{Conclusions}
\label{Sec:conclu}

We have shown that sound  of  a water stream with varying flowrate hitting a pool of water can be successfully used to predict the flowrate across time and hence reconstruct the flowrate trajectory of the varying flow. This is the first study to simultaneously test two approaches against the same real-world measurement. Of the two approaches investigated, the ML method offers good agreement and can robustly predict absolute flowrate, however, with a slight computational overhead. The acoustic parameter method is quick and computationally straight forward; it yields only the relative flowrate but nevertheless still offers good estimation of flowrate trajectories. Collecting sound related to varying flow hitting surfaces offers a viable option to determine flowrate and flowrate trajectory. This provides an indirect and non-invasive method to observe liquid-liquid interactions which in turn can reveal key aspects of the flow interactions and other associated physical properties of the flow system.

\section*{Acknowledgement}

This work was funded by the SingHealth Surgery ACP-SUTD Technology and
Design Multi-Disciplinary Development Programme (grant No. TDMD-2016-1). We also thank our volunteers who help us in collecting data.

\bibliographystyle{IEEEtranN}
\bibliography{mybib}

\end{document}